\documentclass[aps,prl,groupedaddress,a4paper,twocolumn,showpacs]{revtex4}

\usepackage{graphicx}

\begin{document}

\title{Noise-seeded spatio-temporal modulation instability in normal dispersion}

\author{D. Salerno,$^{1,2}$ O. Jedrkiewicz,$^1$ J. Trull,$^3$  G.
Valiulis,$^4$, Antonio Picozzi$^5$ and P. Di Trapani$^1$}

\affiliation{$^1$ INFM e dipartimento di fisica e matematica,
Universit\'a dell'Insubria, Via Valleggio 11, 22100 Como, Italy.\\
$^2$Facolt\'a of Scienze,
Universit\'a di Milano, Via Celoria 16, 20132 Milano, Italy.\\
$^3$Universitat Politecnica de Catalunya. Departament de Fisica i
Enginyeria Nuclear C/ Colom 1, 08222 Terrassa (Barcelona) Spain
\\
$^4$ Department of Quantum Electronics, Vilnius University,
Sauletekio 9-2040 Vilnius, Lithuania.\\
$^5$ CNRS, Laboratoire de Physique de la Matière Condensée,
Universit\'e de Nice, France }

\date{\today}

\begin{abstract}
In optical second harmonic generation with normal dispersion, the
virtually infinite bandwidth of the unbounded, hyperbolic,
modulational instability leads to quenching of spatial
multi-soliton formation and to the occurrence of a catastrophic
spatio-temporal break-up when an extended beam is let to interact
with an extremely weak external noise with coherence time much
shorter than that of the pump.
\end{abstract}


\maketitle

The noise-seeded instability of extended  wave packets (WP) in
conservative evolutional non-linear systems
is a general and relevant phenomenon in wave physics, whose main
manifestations are the appearance of regular modulations, wave
break up and, eventually, localized or soliton-like substructures.
The theory used for the description and interpretation of the
resulting rich phenomenology is that of the modulational
instability (MI) of plane and monochromatic waves, extensively
introduced  in the context of gravity waves in deep waters
\cite{Benjamin} and applied for several different systems
including plasmas \cite{Kakutami}, electric circuits
\cite{Bilbault}, Bose-Einstein condensate \cite{Khawaja} and, of
course, optics \cite{Bespalov}. The usually addressed MI signature
is the \emph{preferential noise amplification at a given,
intensity dependent, (spatial or temporal) frequency}, which
causes regular modulation in the direct space and side-bands in
the spectral domain. This feature suitably describes the MI of
both mono-dimensional (1D) systems and of multi-dimensional
"elliptical" ones, \emph{i.e.} those supporting equi-sign linear
phase modulation in all the available dimensions. However, it is
generally not adequate for "hyperbolic" systems, where opposite
signs occur for different dimensions. The elliptical is the most
frequently encountered regime in case of matter waves in isotropic
media. The hyperbolic, in contrast, is the typical case of optical
WPs in normally dispersive bulk media, diffraction and chromatic
dispersion leading in this case to linear phase modulations with
opposite signs. Recently, dispersion-management techniques based
on use of periodic potentials \cite{Jaksch} have made the
hyperbolic regime of great interest also for the Bose-Einstein
Condensate (BEC) waves.
\begin{figure*}\begin{center}
\includegraphics[width=16.cm]{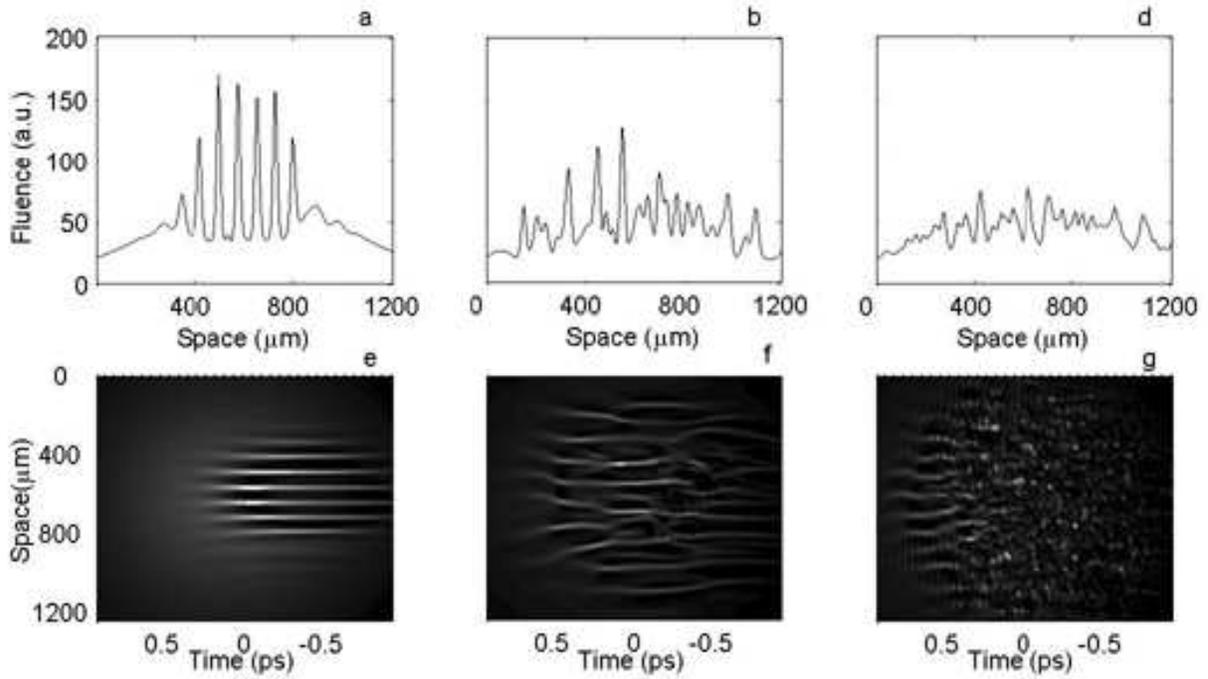}
\caption{Calculated space-time profiles (bottom figure) and
corresponding integrated fluence profiles (top figure) for the
fundamental harmonic propagated in a $50 {\rm mm}$ lithium
triborate (LBO) crystal in regime of $\Delta
k=2k(\omega_{0})-k(2\omega_{0})=5 cm^{-1}$, in absence of noise (a
and d), and with two different kinds of noise (b and e): $6{\rm
nm}$ and $25{\rm mrad}$ of bandwidth, (c and f): $20{\rm nm}$ and
$50{\rm mrad}$. The intensity level of noise is $1\%$. The input
is a gaussian FH WP with $1{\rm ps}$ FWHM duration, $600{\rm\mu m}
$ FWHM beam width, and $40{\rm GW/cm}^2$ peak intensity.
Calculations are performed in the frame of 1D (spatial) + 1D
(temporal) + 1D (propagation) model. Asymmetry in the temporal
coordinate is due to group-velocity mismatch.} \label{fig1}
\end{center}
\end{figure*}
\newline\indent
As clearly pointed out in the first theoretical analysis of
hyperbolic MI performed by Luther et Al. \cite{Luther}  for Kerr
non-linearity in optics, and also evident from the analysis of the
$X^{(2)}$-driven MI \cite{Trillo}, the the key feature that
distinguishes the hyperbolic (normal) from the elliptic
(anomalous) instability regime is that \emph{the MI gain profile
in the k-$\omega$ space is unbounded} in the first (and only in
the first) case. Indeed, in the frame of the usually adopted
parabolic approximation for the material dispersion, we should say
that \emph{any fluctuation with arbitrarily large spatial and
temporal frequency shift respect to the carrier mode has to be
amplified, provided that both shifts lye on the suitable
hyperbolic surface in the k-$\omega$ domain}.
\newline\indent
The unbounded feature of the hyperbolic MI rises two relevant
questions: (i) the first one concerns the interpretation of the
number of studies reguarding \emph{spatial} MI in
multi-dimensional systems. Indeed, following the first
experimental MI demonstration in (anomalous dispersive)
1D-temporal optical fibers \cite{Tai}, a number of experiments
have been performed in multi-dimensional (\emph{i.e.} in planar
wave guides or in bulk samples) normal-dispersion materials,
addressing the 1D \emph{spatial} break up of extended beams driven
by quadratic \cite{Fuerst, Fang} as well as cubic
\cite{Malendevich} ultrafast non-linear response. Surprisingly
enough, the results were successfully interpreted in terms of the
direct spatial analogous of the temporal MI of above, as if the
temporal degree of freedom and so the hyperbolic nature of the
instability had not taken any part in the process. (ii) The second
question concerns the scenario that one should expect when the
system interacts with a very broad band (\emph{i.e.} virtually
$\delta$-correlated) noise. In this case, in fact, the unbounded
feature of the MI gain should lead one to forecast a catastrophic
break in the ST domain (\emph{e.g.} down to the numerical grid in
calculations, in the quoted approximation), no matter how weak the
input noise is. We note that the possible occurrence of such
catastrophic dynamics has never been considered in the literature.
In fact, hyperbolic MI has been studied only for the case of
bell-shaped, noise-free, input wave packets, \emph{e.g.} for
investigating the impact of ST self-phase-modulation (SPM) on
filament formation, pulse splitting and related phenomena
\cite{Luther}. To the best of our knowledge, the role of the noise
has been considered only in the context of 1D models. The aim of
this work is that of providing experimental evidence of the
genuine hyperbolic feature of the \emph{noise-seeded} MI in bulk,
normally dispersive, optical systems.
To this end we performed experiments and
calculations in which, for the first time to our knowledge, a
controlled, broad-band noise is injected together with the strong,
(quasi) plane and monochromatic pump into the system.
\begin{figure*}[htb]\begin{center}
 \includegraphics[width=16.cm]{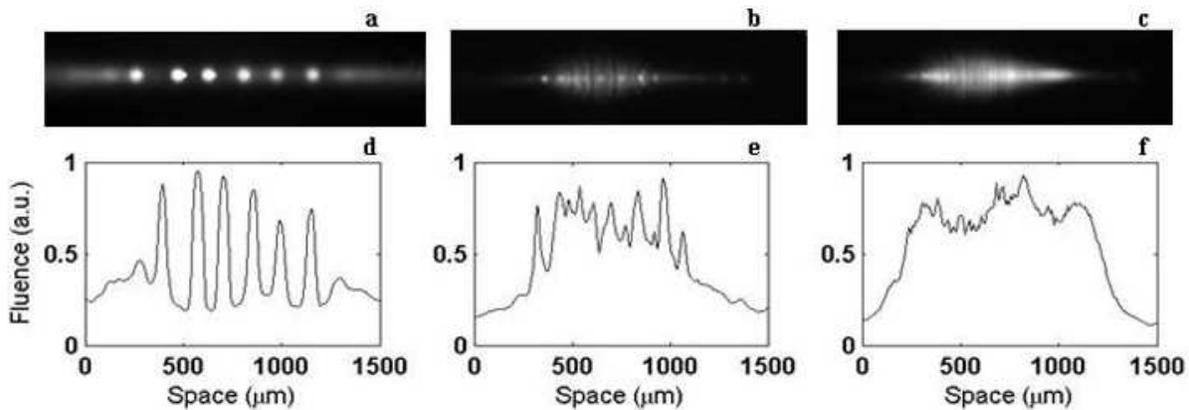}
   \caption{Measured fluence profiles (top figure) of FH WP recorded by a
   CCD camera at the output of the $50{\rm mm}$-LBO crystal,
and (bottom figure) corresponding beam profiles (note that the
horizontal scales of images and profiles are not the same) in
conditions: (a) and (d), with no noise; (b) and (e) with noise
intensity $0.01\%$ of that of the pump; (c) and (f), with $0.1\%$
of noise. The noise spatial and temporal BWs were respectively of
$60{\rm mrad}$ and $100{\rm nm}$. The input pump intensity was
$20{\rm GW/cm}^2$.} \label{fig.2}.
\end{center}
\end{figure*}
\newline\indent
The particular system that we have chosen to investigate is the
same as in Ref. \cite{Salerno}, \emph{i.e.} that of an optical WP
shaped as a large (with respect to diffraction), elongated beam
and long (with respect to dispersion) pulse that propagates in a
$X^{(2)}$ non-linear crystal tuned for second harmonic (SH)
generation close to phase matching. For the chosen crystal
(lithium triborate, LBO) and wavelength (first harmonic, FH,
1055nm) the chromatic dispersion (see caption to Fig. 1 for
details) is such that one should expect MI to take place in the
hyperbolic regime \cite{Trillo}. We performed numerical
calculations in the frame of 2D+1 model integrating (\emph{via}
FFT, split step and Runge Kutta algorithms, with up to 15fs $6{\rm
\mu m}$ grid and $40{\rm \mu m}$ step) the  $\chi^{(2)}$
coupled-wave equations for FH and SH envelopes $E_{j}(z,x,t)$
\begin{equation}\label{SHGeqs}
\begin{array}{l}
{\displaystyle \hat{L}(\omega_0)~E_1 + \chi E_2 E_1^* \exp(-i
\Delta
k z) =  0,} \\
{\displaystyle \hat{L}(2 \omega_0)~E_2 + i \delta V \partial_t E_2
+ \chi E_1^2 \exp(i\Delta k z) = 0,}
\end{array}
\end{equation}
where $\hat{L}(\omega) \equiv i \partial_{z} + \left( 2k(\omega)
\right)^{-1} \partial^2_{xx} - (k''(\omega)/2)
\partial_{tt}^2$,
$k''$ is the group-velocity dispersion (GVD),  $\delta
V=k'(2\omega_{0})-k'(\omega_{0})$ weighs the group-velocity
mismatch (GVM), and $\Delta k=2k(\omega_{0})-k(2\omega_{0})$.
Figure 1 gives the calculated fluence (\emph{e.g.} energy density)
profile of the FH at the output of a 50 mm crystal (top), together
with the corresponding spatio-temporal (ST) intensity maps
(bottom). Figs. 1a,d, which refer to noise-free input, show the
occurrence of a regular, highly contrasted, spatial break up of
the beam into a spatial-soliton array, which appears as the
consequence of MI seeding by the deterministic wave-envelope
modulation (WEM \cite{Salerno}).
When the ST noise is injected, WEM and noise-seeded MI compete and
the results are those shown in Fig. 1b,c, and Fig.1e,f, where the
input-noise bandwidth (BW) is increased (from left to right) while
keeping fixed the noise intensity at $1\%$ of the level of the
pump. See
how the impact of the noise dramatically increases on enlarging
its BW, owing to the unbounded instability. Note how the noise,
instead of deepening the spatial modulation (as it occurs in the
frame of the 1D+1 models \cite{Fuerst}), quenches it almost
completely. The reason is the appearance of a "chaotic gas" of
localized ST structures (Fig. 1f), which gets washed out by the
temporal integration.
\newline\indent
In order to verify if the outlined catastrophic behavior is a
genuine physical effect or an artefact of the approximations
adopted we performed a SH-generation laboratory experiment in
similar conditions to those which Fig. 1 refers to. To this end we
used a strongly elongated (1000x70 $\mu m$), long (1ps) pump WP,
as clean as possible from any spatial or temporal substructure,
provided by a CPA Nd:Glass laser (TWINKLE, Light Conversion). Then
we superimposed to the pump a weak, broad-bandwidth, ST noise, of
controllable intensity, generated on a separate channel by a
broad-band quantum-noise parametric amplifier. For the noise
generation we used a 15mm LBO crystal pumped by the SH of (a
portion of) our pump pulse. Both pump and noise WPs were launched
synchronously into a 50 mm LBO crystal, tuned for phase matched SH
generation. The spatial and temporal BWs of the noise field at the
input of the SH generator were 100 nm and 60 rad, respectively.
Fig. 2 (top) shows the fluence distributions of the FH beam at the
crystal output facet as recorded by a CCD camera and suitable
imaging optics. The corresponding profiles (along the long axis of
the beam) are given in Fig. 2 (bottom) for a more quantitative
description. The results in the left, center and right part of the
figure refer to average noise fluence 0, 0.01$\%$ and 0.1$\%$ of
that of the pump, respectively, for a fixed noise ST BW. The
resulting scenario fully confirms the model prediction. Indeed the
noise-induced quenching of the WEM-seeded spatial MI takes place
in the experiment for a lower noise level than in calculations,
which indicates that the accessible BW is even larger than the
computational BW used for obtaining the Fig. 1c,f results.
\newline\indent
The described, \emph{near-field}, measurements have the obvious
limitation of confirming the model prediction only on the basis of
a time-integrated effect, the underlying ST structure not being
detectable by any technique. Moreover, both the numerical and the
experimental results that we have presented do not produce a
direct evidence of the hyperbolic nature of instability. In what
follows, in order to overcome these limitations, we illustrate the
complementary, \emph{far-field} analysis, concerning the
characterization of the angular spectra (AS) of the generated
field. Figure 3a contains the calculated AS (\emph{i.e.} the
square modulus of the field Fourier transform) of a FH profile
analogous to that in Fig. 1f (see the caption for details). The
bright central spot corresponds to the spectrum of the input pulse
while the surrounding structure describes the amplified
fluctuations. Note the evident hyperbolic shape of the instability
region, which coincides with the region where the calculated
MI-gain profile is the largest \cite{Trillo}. We verified that, no
matter the size of the Fourier space, the instability always
reaches the border of the spectral box (in case of large enough
noise BW), thus confirming the unbounded nature of the MI process
in the frame of the adopted model. For the far-field experiment,
we measured the FH AS by placing the entrance slit of a
large-numerical-aperture imaging spectrometer at the focal plane
of a positive lens. The detected AS in the short-wavelength branch
of the spectrum (the long-wavelength one is not accessible due to
sensitivity cut off of the silicon CCD detector) is reported in
Fig. 3c. The detected portion of the AS (the radiation at large
angles was clipped by the very narrow aperture of our 3x3x50
mm$^3$ LBO crystal) exhibits a good qualitative agreement with
calculations (see the zoom in fig. 3b) and confirms the genuine
hyperbolic feature of the instability process.
\begin{figure}[htb]\begin{center}
 \includegraphics[width=8.5cm]{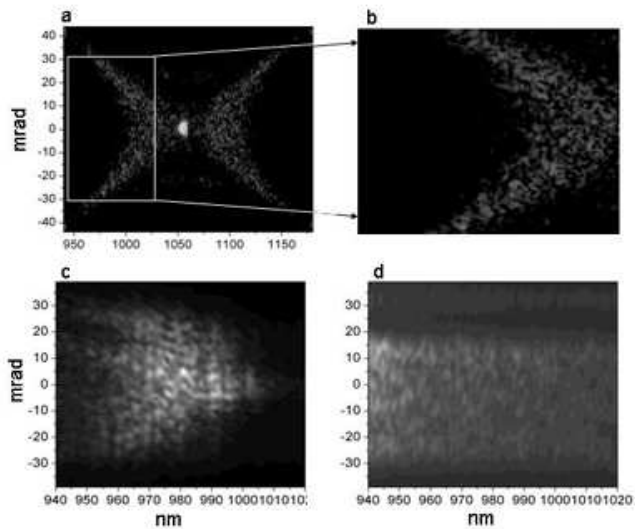}
   \caption {a) Calculated angular spectra (AS) of the FH field for operating conditions
   as those in Fig. 1f, but for larger input-noise BW (namely: $40{\rm nm}$ and
$50{\rm mrad})$; (b) Zoom of (a) corresponding to the region of
detection; (c) measured AS of the radiation exiting the crystal;
(d) measured AS of the input noise. Grey colors refer to
logarithmic scale.} \label{fig.3}
\end{center}
\end{figure}
\newline\indent
In conclusions, in regime of second-harmonic generation \emph{with
normal dispersion}, by superimposing on a intense, clean, pump
wave packet and a weak spatio-temporal noise (with coherence time
much shorter than that of the pump) we have shown that the
noise-seeded MI develops in the spatio-temporal domain. Due to the
unbounded feature of the instability, and so the large response of
the system to white noise, a noise as weak as 0.01$\%$ of the pump
leads to quenching of the spatial deterministic beam break up and
spatial-soliton formation caused by wave envelope modulation
(WEM). The analysis of the results in the spectral domain outlined
the genuine hyperbolic feature of the instability, which couples
different frequencies to different angles. Because of this
coupling, the quenching of detectable spatial effects cannot be
simply interpreted as the averaging of several independent spatial
structures, occurring for different time slices of the wave
packet. As pointed out in \cite{Fang}, the typical (amplitude and
phase) fluctuations that were triggering the instability in
previous experiments (without external noise injection) were
probably caused by laser-beam, or optical-component or non-linear
sample imperfections, thus leading to a "frozen-noise" with the
same coherence time of the pump (that virtually coincides with the
pulse duration, all lasers operating close to the transform
limit). This might explain why the resulting instability was so
successfully described in the frame of monochromatic models.
Finally, we expect that the described, hyperbolic instability
should play a dramatic role when the field is strong enough to
probe vacuum state fluctuations, which indeed provide the source
of a virtually $\delta$-correlated noise. We expect that
quantum-noise seeded hyperbolic instability should dominate not
only the $X^{(2)}$ parametric-amplification regime (as evident,
for example, in ref. \cite{Devaux} and also in the more recent
ref. \cite {Zeng}) but also the classical, unseeded,
second-harmonic generation process and, possibly, the Kerr regime
too. This might explain the "spontaneous" quenching of spatial
break-up seen at high pumping in \cite{Salerno, Fang}. Owing to
the unbounded nature of the instability, the robustness of
non-linear dynamics of normally dispersive media with respect to
the interaction with the quantum noise represents therefore a
crucial issue with deserves further investigation.

The Authors acknowledge support from FIRB 2001 (Italy), DGICYT
BFM2002-04369-C04-03 (Spain), contracts.

\end{document}